\documentclass[12pt]{article}
\usepackage[hmargin=2.5cm,vmargin=3cm]{geometry}
 \usepackage{graphicx}
 \usepackage{amsmath}
 \usepackage{amssymb}
  \usepackage{enumerate}
\usepackage{cite}

\begin{document} \title{Time of arrival and localization of relativistic particles} \author {Charis Anastopoulos\footnote{anastop@physics.upatras.gr}   and   Ntina Savvidou\footnote{ksavvidou@upatras.gr}\\
 {\small Department of Physics, University of Patras, 26500 Greece} }

\maketitle

\begin{abstract}
Constructing observables that describe the localization of relativistic particles is an important foundational problem in relativistic quantum field theory (QFT). The description of localization in terms of single-time observables leads to conflict with the requirement of causality. In this paper, we describe particle localization in terms of time-of-arrival observables, defined in terms of the interaction between  a quantum field and a measuring apparatus. The resulting probabilities are linear functionals of QFT correlation functions. Specializing to the case of a scalar field, we identify several time-of-arrival observables differing on the way that the apparatus localizes particle-detection records.  Maximum localization is obtained for a unique observable that is related to the Newton-Wigner position operator. Finally, we define a measure of localizability for relativistic particles in terms of a novel time-energy uncertainty relation for the variance of the time of arrival.

\end{abstract}

\section{Introduction}
\subsection{The relativistic localization problem}
Relativistic Quantum Field Theory (QFT) is characterized by an intricate interplay between particles and fields. While the dynamics of the theory are expressed in terms of fields, in experiments we observe particles \cite{Ana08}. However, the description of the latter  is a source of  conceptual problems.  Particles have two key properties that are crucial to the physical interpretation of experiments. First, they are discrete, countable entities: the number of particle-detection events  is always a positive integer. Second, they are localizable entities: every particle-detection event is localized in space and in time.

Particle discreteness is elegantly expressed in terms of the QFT spacetime symmetries. It follows from the requirement that the mass operator $\hat{M}^2 = \hat{P}^{\mu}\hat{P}_{\mu}$---defined in terms of the generators $\hat{P}^{\mu}$ of spacetime translations---has discrete spectrum \cite{Haag}. Hence, the corresponding subspace contains single-particle states that correspond to unitary irreducible representations of the Poincar\'e group.

  Unlike discreteness,  localization is highly problematic. Existing definitions of localizing observables  conflict   the requirement of relativistic causality, as evidenced by several theorems \cite{Schl71, Heg98, Mal96}. Assume, for example, that localization is defined with respect to some spatial observable ${\pmb x}$, leading to  a sufficiently localized probability distribution $\rho({\pmb x}, t)$ for ${\pmb x}$ at some moment of time $t$. Then, time evolution  leads to a probability distribution $\rho({\pmb x}, t')$ that evolves superluminally at latter times $t'$.

The localization problem originates   from the fact that particle trajectories are not fundamental concepts in quantum theory. In absence of trajectories, localization is usually expressed in terms of single-time position observables. The most well-known example is the Newton-Wigner position operator $\hat{\pmb x}$ \cite{NW46, Wight62}. There are three problems in the physical interpretation of such observables. First, they do not transform covariantly under Lorentz transformations. This is due to the fact that there exists no time operator in quantum theory  \cite{Pauli};  the position observable is not the spatial component of some covariant four-vector.   Second, the associated   probability distributions exhibit superluminal behavior \cite{RoUs87}. Third, massless particles with spin $s \neq 0$  (e.g., photons) are not localizable \cite{Wight62}, in contrast to the existence of well-localized  photon-detection records.

Hence, even if there are many ways to construct densities $\rho({\pmb x}, t)$ with respect to some position variable ${\pmb x}$, we cannot interpret  $\rho({\pmb x}, t)d^3x$  as the probability to find the particle in the volume $d^3x$ at the space point ${\pmb x}$. As a result,  there is no obvious measure to quantify the localization of a quantum state,  like, for example, the position variance in non-relativistic quantum mechanics.

The localization problem is not an artefact of the particle description, it persists even in a full QFT treatment. The definition of appropriate  observables for localized measurement records is necessary for the conceptual completeness of relativistic quantum theory, irrespective of whether one employs particle or field degrees of freedom. It has been
 recognized that localization observables should not be viewed as attributes of particles (or even of their associated  fields),
  but as attributes of the interaction between particles (or fields) and a measuring apparatus \cite{Haag, PeTe}. In this perspective, a solution to the localization problem requires
 a consistent generalization of quantum measurement theory to   relativistic QFTs.

 \subsection{Localization via time-of-arrival observables}

The use of single-time position observables in relativistic systems suffers from two major conceptual problems.   First, a position measurement at fixed time $t$ corresponds to an {\em instantaneous} scan of all space in order to determine the particle's position.  It is implausible that such a  process can be expressed in terms of a localized field-apparatus interaction. A breakdown of causality in the measurement of such observables is not a surprise.

  Second, we do not measure single time observables in the laboratory.   Actual particle detectors (e.g., photographic plates, silicon strips) have finite extension, and they are made sensitive for a long time interval during which particles are detected. This means that the location of a detection event is a fixed parameter of the experiment;  the actual random variable is the detection time. Hence, it is more realistic to express particle localization  in terms of {\em time-of-arrival  measurements} rather than position ones \cite{Wer86}.

The simplest example of a time-of-arrival measurement is the following.
 A particle is prepared on an initial state $|\psi_0 \rangle$ that is localized around $x = 0$ and has positive mean momentum. If a  detector is placed at $x = L$, what is the  probability  $P(L, t)dt$ that the particle is detected at $x = L$ at some moment between $t$ and $t+\delta t$?   The lack of a self-adjoint operator for time  means that we cannot employ Born's rule in order to obtain an unambiguous answer. In spite of the problem's apparent simplicity, no unique time-of-arrival probability exists.
  Several different approaches   have been developed, each with a different conceptual framework. For reviews on the time-of-arrival issue (mainly in non-relativistic quantum mechanics), see, Refs.  \cite{ML, ToAbooks}.

In this paper, we   construct relativistic time-of-arrival observables that describe particle localization using the
 Quantum Temporal Probabilities (QTP) method  \cite{AnSav12}.  QTP was first developed in order to address the time-of-arrival problem  \cite{AnSav12, AnSav06}. It was then extended to  a general algorithm for constructing quantum probabilities for temporally extended observables, i.e., observables that are not defined at a single pre-determined moment of time.  QTP  has  been applied to the temporal characterization of tunneling  \cite{AnSav13, AnSav08, AnSav17b}, to non-exponential decays \cite{An08}, to the response and correlations of accelerated particle detectors \cite{AnSav11}, and for defining entanglement witnesses in terms of temporal observables  \cite{AnSav17}.

The key idea  is to distinguish between the time parameter of Schr\"odinger equation from the time variable associated to  particle detection \cite{Sav99, Sav10}. The latter are then treated as   macroscopic quasi-classical variables associated to the detector degrees of freedom. Here, we use the word `quasi-classical' as in the decoherent histories approach to quantum theory \cite{Gri, Omn1, Omn2, GeHa, hartlelo}. It refers to coarse-grained quantum variables that satisfy appropriate decoherence conditions, and they approximately satisfy  classical evolution equations  \cite{GeHa, hartlelo}. Hence, although  the detector is described  in microscopic scales by quantum theory, its macroscopic records are expressed  in terms of classical spacetime coordinates.

\subsection{Results}
In our approach, localization is expressed in terms of  time-of-arrival observables, defined as emergent attributes of the interaction between a quantum field and a measuring apparatus---see Refs.    \cite{ ECM08, HaYe09, HELM} for  non-relativistic measurement models for the time-of-arrival. We make no reference to particles at the fundamental level.
Time-of-arrival probabilities are expressed in terms of QFT correlation functions, and the particle description {\em emerges} from the structure of the correlation functions. This is particularly important for the broader applicability of the method.  Most experimentally accessible quantum states do not have definite particle number: there exist multiple decay channels and the creation of soft photons must always be taken into account \cite{WeinbergQFT}.

The main results of this paper are the following. First, we derive a general formula for the particle detection probability $P(X) d^4X$, for detection within a four-volume $d^4X$ at a spacetime point $X$.
We emphasize that $P(X)$ is a {\em genuine density} with respect to $X$. This is important because the  probability density  given by Born's rule $P({\pmb x}, t) = |\psi({\pmb x},t)|^2$ is a density with respect to ${\pmb x}$, but not with respect to $t$, i.e.,  $t$ appears as a parameter and not as a random variable. In contrast, time appears as a genuine random variable in the probability densities derived here.

  The probability density $P(X)$   is a linear functional of a QFT two-point function that characterizes the field-detector coupling. It turns out that all information about the apparatus is contained in a single function that describes the four-momentum content of the apparatus pointer state.
 This probability formula is completely general: it applies to all QFTs (free and interacting) and to all types of particle.  It is a significant improvement over earlier QTP formulas   \cite{AnSav12}, in terms of generality and rigor.

Second, we analyze in detail the case of a free scalar field. We construct a general  class of Positive-Operator-Valued Measures (POVM) for the time of arrival of a relativistic particle with mass $m$ and spin $s = 0$.  Each POVM  is uniquely characterized by a {\em localization operator} $\hat{L}$ that describes how the  apparatus localizes particle detection events.
 Maximum localization is achieved for a specific  $\hat{L}$ that is closely related to the Newton-Wigner position operator \cite{NW46}. In the non-relativistic limit, the maximum-localization POVM reduces to the one of Kijowski \cite{Kij}.

Third, we derive a new time-energy uncertainty relation in the form of a lower bound to the deviation  $\Delta t$ of the time-of-arrival,
\begin{eqnarray}
(\Delta t)^2 \geq \frac{1}{4(\Delta H)^2} +\frac{m^2}{4} \langle \hat{H}^{-2}\hat{p}^{-4}\rangle; \label{1}
\end{eqnarray}
 $(\Delta H)^2$ is the variance of the Hamiltonian $\hat{H} = \sqrt{\hat{p}^2+m^2}$ and $\hat{p}$ is the standard momentum operator.
 This uncertainty relation is valid for all  states with strictly positive momentum, and it does not depend on the properties of the detector. We argue that the r.h.s of Eq. (\ref{1}) is a measure for the spacetime localization of  relativistic particles.

\medskip

We
describe  the interaction between detector and microscopic particles in terms of a Hamiltonian that is a local functional of quantum fields. This implies that the field correlation functions that appear in the probability assignment  satisfy  QFT causality conditions.
For this reason, we strongly believe that the time-of-arrival observables constructed here are fully consistent with relativistic causality. Further work is required in order to substantiate this claim. It is necessary to extend the results of this paper to set-ups that involve multiple, spatially separated detectors---see, Ref. \cite{AnSav17} for the non-relativistic case---and show that the resulting probabilities do not
lead to superluminal propagation of information.

In this paper, we focus  on the localization of particles with zero spin.   However, our approach can be straightforwardly generalized to particles of any spin. We believe that such a generalization will allow us to quantify localization even in the   challenging cases of massless particles with spin (photons, gravitons), and of  particles in mass-eigenstate superpositions (neutrinos).

\medskip

The structure of the paper is the following. In Sec. 2, we derive the probability formula for particle detection that applies to any QFT. In Sec. 3, we specialize this formula to the case of free scalar fields. In Sec. 4, we construct   time-of-arrival probability measures for spinless relativistic particles, and we analyze the properties of the localization operator. In Sec. 5, we derive an uncertainty relation for the time of arrival. In Sec. 6, we summarize and discuss our results.

\section{Relativistic covariant measurement model }

\subsection{The QTP probability formula}
In this section, we present the main probability formula, to be employed in our detection model. For a derivation, see Refs. \cite{AnSav12, AnSav13, AnSav17}, and also the appendix. This construction originates from the treatment of spacetime coarse-grainings in the decoherent histories approach to quantum mechanics \cite{Har91, hartlelo}.

We consider a composite physical system that consists of a microscopic and a macroscopic component. The microscopic component is the quantum system to be measured and the macroscopic component is the measuring device. We denote the Hilbert space associated to the composite system
 by
   ${\cal H}$ and the associated Hamiltonian by $\hat{H}$.

   We describe a measurement event as a transition between two complementary subspaces of ${\cal H}$. Hence, we consider a splitting of
 ${\cal H}$  into two subspaces: ${\cal H} = {\cal
H}_+ \oplus {\cal H}_-$. The subspace ${\cal H}_+$ describes the accessible states of the system given that the event under consideration is realized. For example, if the event is a detection of a microscopic particle by  an  apparatus,   the subspace ${\cal H}_+$ corresponds to all states of the apparatus compatible with the macroscopic record of detection.  We
denote  the projection operator onto ${\cal H}_+$ as $\hat{P}$ and the projector onto ${\cal H}_-$ as $\hat{Q} := 1  - \hat{P}$.

We construct the probability density with respect to time that is associated to the transition of the system from ${\cal H}_-$ to ${\cal H}_+$. We consider transitions that are correlated with the emergence of a macroscopic record of observation. Such transitions are {\em logically
 irreversible}. Once they occur, and a measurement outcome has been recorded,   further time evolution of the system does not affect our knowledge that they occurred.

After the transition has occurred,  a pointer variable $\lambda$ of the measurement apparatus takes a definite value. Let
$\hat{\Pi}(\lambda)$ be  positive operators that correspond to the different values of $\lambda$.  For example,
when considering transitions associated with particle detection, the projectors $\hat{\Pi}(\lambda)$  may be correlated  to the position, or to the momentum of the microscopic particle. Since $\lambda$ has a value only under the assumption that a detection event has occurred, the alternatives
  $\hat{\Pi}(\lambda)$ span the subspace
  ${\cal H}_+$ and not the full Hilbert space ${\cal H}$. Hence,
  \begin{eqnarray}
  \sum_\lambda \hat{\Pi}(\lambda) =
\hat{P}.
\end{eqnarray}
Since $\lambda$ refers to a {\em macroscopically distinguishable record} on a measurement apparatus, the observables  $\hat{\Pi}(\lambda)$ are highly coarse-grained. We will denote the sampling width of $\lambda$ by    $\sigma_{\lambda}$.

We assume an initial state $|\psi_0\rangle \in {\cal H}_+$, and a Hamiltonian of the form $\hat{H} = \hat{H}_0 + \hat{H}_I$, where $[\hat{H}_0, \hat{P}] = 0$ and $\hat{H}_I$ is a small perturbation. This means that the transition from ${\cal H}_-$ to ${\cal H}_+$ is effected only by the interaction Hamiltonian  $\hat{H}_I$.

With the above assumptions, we derive the probability density  $P(\lambda, t)$  for a transition at time $t$ that leads to a value $\lambda$ for the measured observable
\begin{eqnarray}
 P(\lambda, t) = \int ds ds' \sqrt{f(t-s) f(t-s')} Tr \left[\hat{C}(\lambda, s)
 \hat{\rho}_0 \hat{C}^{\dagger}(\lambda, s')\right] \label{ampl6}
\end{eqnarray}
where $\hat{\rho}_0 =|\psi_0\rangle \langle \psi_0|$, and
\begin{eqnarray}
 \hat{C}(\lambda, t) = e^{i \hat{H}_0t} \sqrt{\hat{\Pi}}(\lambda) \hat{H}_I e^{-i \hat{H}_0t},
\label{perturbed}
\end{eqnarray}
 to leading order in the perturbation $\hat{H}_I$.
 The positive functions $f(s)$ describe   sampling for the transition time, with a sampling width $\sigma_T$. They
 are localized around $s = 0$, and they are normalized: $\int ds f(s) = 1$.
The function  $P(\lambda, t)$ is a  {\em density} with respect to both $\lambda$ and $t$, i.e., it defines genuine probability distributions with respect to the time $t$ of transition.

 The derivation of Eq. (\ref{ampl6}) requires a {\em decoherence condition}
\begin{eqnarray}
Tr \left[\hat{C}(\lambda', s')
 \hat{\rho}_0 \hat{C}^{\dagger}(\lambda, s) \right] \simeq 0, \label{deccond}
\end{eqnarray}
if either $|s - s'| > \sigma_T $, or $|\lambda - \lambda'| > \sigma_{\lambda}$. The decoherence condition is necessary for the existence of records of observation. It holds  for any system that contains   a macroscopic component such as a measuring apparatus  \cite{GeHa, hartlelo}.

 The probability density (\ref{ampl6})  is of the form
 $Tr[\hat{\rho}_0 \hat{\Pi}(\lambda, t)]$,
where
\begin{eqnarray}
\hat{\Pi}(\lambda, t) = \int ds ds' \sqrt{f(t - s) f(t - s')} \hat{C}^{\dagger}(\lambda, s') \hat{C}(\lambda, s). \label{povm2}
\end{eqnarray}

 The operator $\sum_{\lambda} \int_0^{\infty} dt  \hat{\Pi} (\lambda, t)$ corresponds to the total probability that an event has been recorded in the time interval $[0, \infty)$. Consequently, the operator
\begin{eqnarray}
\hat{\Pi} _{\emptyset} = \hat{1} -  \sum_{\lambda} \int_0^{\infty} dt \hat{\Pi} (\lambda, t), \label{nodet}
\end{eqnarray}
 corresponds to the alternative $ \emptyset$ that no detection took place. The lack of a measurement record may be due to the fact that some of the particles in the statistical ensemble "missed" the detector, or due to the  non-zero probability that the interaction of the microscopic  particles with the apparatus leaves no record.

In the systems considered here,
 $\hat{\Pi} _{\emptyset} $ is always positive. Hence,  then $\hat{\Pi} _{\emptyset} $ together with the positive  operators Eq. (\ref{povm2})
 define a   POVM that is associated to a complete set of alternatives. These POVMs define {\em time-of-arrival observables}.

\subsection{A quantum field detector}
Next, we apply Eq. (\ref{ampl6}) to the measurement of a quantum field by a single measuring apparatus. First, we identify all mathematical objects that enter into the probability formula (\ref{ampl6}).

 \bigskip

\noindent {\em Hilbert space structure.}
The quantum field is defined on a  Hilbert space ${\cal F}$. We denote the   Heisenberg-picture field operators as $\hat{\Phi}_r(X) := \hat{\Phi}_r({\bf x}, t)$, where
$r$ is a collective index that may include both spacetime and internal indices. The fields $\hat{\Phi}_r(X)$ may include both bosonic and fermionic components, and they may be either free or interacting.

In a relativistic system, the Hilbert space ${\cal F}$ carries a unitary representation of the Poincar\'e group. A unitary operator $\hat{U}(\Lambda, a)$ is associated to each element $(\Lambda, a)$  of the  Poincar\'e group, so that the fields $\hat{\Phi}_a(X)$ transform as

\begin{eqnarray}
\hat{\Phi}_r(X) \rightarrow \hat{U}^{\dagger}(\Lambda, a) \hat{\Phi}_r(X) \hat{U}(\Lambda, a) = D_{r}^{r'}(\Lambda) \hat{\Phi}_{r'}(\Lambda^{-1} X - a),  \label{poincare1}
\end{eqnarray}
for some matrix $D_r^{r'}(\Lambda)$.

 The measuring apparatus is described by a Hilbert space ${\cal K}$. A detection event is associated to a transition between two complementary subspaces of ${\cal K}$, i.e.,
${\cal K} = {\cal K}^- \oplus
{\cal K}^+$. The subspace  ${\cal K}^-$   corresponds to the absence and  the subspace ${\cal K}^+$  to the presence of a macroscopic measurement record.  
  We denote  the projector associated to   ${\cal K}^+$ by $\hat{E}$.


The Hilbert space  ${\cal H}$ describing the total system including the quantum fields and the measurement devices is ${\cal H} = {\cal F} \otimes {\cal K}$.

 \bigskip

 \noindent {\em Dynamics.} We assume that the detector is static on a Lorentz frame with time parameter $t$. In this frame, the Hamiltonian $\hat{H}_0$ for the non-interacting system is $\hat{H}_{\phi}\otimes I + I \otimes \hat{\mathfrak{h}}$, where  $\hat{H}_{\phi}$  is the Hamiltonian of the quantum field, and $\hat{\mathfrak{h}}$ the Hamiltonian of the detector.

Relativistic causality    implies that the interaction term $\hat{H}_I$  is   a  local functional of the field operators \cite{WeinbergQFT},

\begin{eqnarray}
\hat{H}_I = \int d^3 x \hat{O}_a({\bf x}) \otimes    \hat{J}^a({\bf x}).
  \label{vterm}
\end{eqnarray}
 where $\hat{O}_a({\pmb  x})$ is a composite operator on ${\cal F}$ that is a local functional of the fields $\hat{\Phi}_r$, and $\hat{J}^a({\bf x})$ are current operators  defined on the Hilbert space ${\cal K}$ of the  detector; $a =1, 2, \ldots n$ is a collective index for the composite operators. We will write $\hat{J}^a({\bf x}) = e^{-i\hat{ \bf \mathfrak{p}}\cdot {\bf x}}\hat{J}^a(0)  e^{i\hat{ \bf \mathfrak{p}}\cdot {\bf x}}$, where $ \hat{ \bf \mathfrak{p}}$ is the generator of space translations on the detector Hilbert space ${\cal K}$.

 \bigskip


\noindent {\em The initial state.}
 We   assume a factorized initial state for the total system\footnote{All models of quantum measurement theory presuppose a factorized initial state between detector and apparatus,
In QFT, a generic state of the system does involve correlations between field and apparatus, because their interaction cannot be switched off. The initial state of the apparatus is "dressed" with vacuum fluctuations of the field, which  induce a renormalization of the physical parameters of the detector. However, in any reasonable modeling of a measurement apparatus, dressing should not affect the  correlation between pointer variables and   microscopic degrees of freedom. Its effect should be included into the noise that characterises the evolution of any coarse-grained observable \cite{GeHa}.
Hence,  the use of factorized initial states should not significantly affect the probabilities associated to measurements. Renormalization will be needed, because, strictly speaking, the vector (\ref{factorizedpsi}) does belong in a same Hilbert space where  a Hamiltonian with  the interaction term (\ref{vterm}) exists.  This is not   necessary in this paper, because we work in the lowest order of perturbation theory. The factorization approximation, Eq. (\ref{factorizedpsi}), may lead to small terms in the probabilities that violate causality, but these  correspond to higher order corrections to the probability formula.}, i.e., a state of the form
\begin{eqnarray}
|\psi_0\rangle \otimes |0\rangle \label{factorizedpsi}
\end{eqnarray}
where $\psi_0 \in {\cal F}$ and $|0\rangle \in {\cal K}_-$.

Physical detectors  have an  energy gap between the non-detection and the detection states; otherwise small environmental perturbations would cause a measurement signal. We assume that the Hamiltonian $\hat{\mathfrak{h}}$ of the apparatus  has a unique ground state $|0\rangle$, and a continuum of states separated by a gap $\Delta$ from $|0\rangle$. Hence, the Hilbert space ${\cal K}_-$ consists only of $|0\rangle$. Without loss of generality, we take  $E_0 $ to be zero. Since $[ \hat{\mathfrak{h}}, \hat{\mathfrak{p}}  ] = 0$,  $|0\rangle$ is also an eigenstate of the momentum $\hat{\mathfrak{p}}$. We choose a Lorentz frame so that   $\hat{\mathfrak{p}}|0 \rangle = 0$.

 \medskip

 \noindent {\em Observables.} We will only consider position observables in this paper. Let $\hat{\pmb x}$ be a self-adjoint operator on ${\cal K}_+$, conjugate to  the momentum $\hat{\mathfrak{p}}$, $[\hat{x}_i, \hat{\mathfrak{p}}_j] = i \delta_{ij}\hat{E}$. Let $|{\pmb x}, j \rangle$ be the  generalized eigenvectors of $\hat{\pmb x}$, where $j$ refers to the remaining degrees of freedom in ${\cal K}_+$.

 We define the position sampling operator
 \begin{eqnarray}
  \hat{F}_{\pmb x} = \sum_j \int d^3 x' g({\pmb x} - {\pmb x}')|{\pmb x}', j\rangle \langle {\pmb x}', j|,
 \end{eqnarray}
 where $g({\pmb x} - {\pmb x}')$ is a sampling function of width $\sigma_X$.  We assume that $\int d^3x g({\pmb x}) = 1$, so that $\int d^3x \hat{F}_{\pmb x} = \hat{E}$. By construction,  $\hat{F}_{\pmb x}$
 transforms covariantly under  space translations,
\begin{eqnarray}
 e^{i\hat{ \bf \mathfrak{p}}\cdot{\bf a}} \hat{F}_{\pmb x}  e^{-i\hat{ \bf \mathfrak{p}}\cdot {\bf a}} = \hat{F}_{{\pmb x} - {\pmb a}}. \label{covar}
\end{eqnarray}
We express the positive operator  $\hat{\Pi}({\pmb x})$ on the Hilbert space ${\cal H}$ of the total system   as $ I \otimes   \hat{F}_{\pmb x}$.

\bigskip

\noindent {\em The probability formula.}
Using the expressions above for the field-detector system, Eq. (\ref{ampl6}) becomes
\begin{eqnarray}
P({\pmb x}, t) = \int ds ds' d^3y d^3 y' R^{ab}_{\lambda, t} ({\pmb y}, s; {\pmb y}', s') Tr \left[ \hat{O}_a({\pmb y}, s)\hat{\rho}_0 \hat{O}^{\dagger}_b({\pmb y}', s')\right], \label{probrel}
\end{eqnarray}

where
\begin{eqnarray}
\hat{O}_a({\pmb x}, s) = e^{i \hat{H}_{\phi}s} \hat{O}_a({\pmb x}) e^{-i \hat{H}_{\phi}s}
\end{eqnarray}
are Heisenberg-picture composite operators, and the kernel

\begin{eqnarray}
R^{ab}_{{\pmb x}, t} ({\pmb y}, s; {\pmb y}', s') =\sqrt{f(t-s)f(t - s')} \langle 0  |   \hat{J}^{b\dagger}({\pmb y}')       \sqrt{F}_{\pmb x}   e^{i \hat{\mathfrak{h}(s-s')}}  \sqrt{F}_{\pmb x} \hat{J}^a({\pmb y})  |0\rangle \label{kernelR}
\end{eqnarray}
depends only on properties of the detector.

By Eq. (\ref{covar}), $\sqrt{F}_{\pmb x} \hat{J}^a({\pmb y})  |0\rangle =  e^{-i\hat{ \bf \mathfrak{p}}\cdot {\bf a}}  \hat{F}_{{\pmb x} - {\pmb a}} \hat{J}^a(0)|0\rangle$. The un-normalized vector
\begin{eqnarray}
|\omega, a  \rangle =  \hat{J}^a(0)|0\rangle
\end{eqnarray}
describes an excitation due to particle detection localized around ${\pmb x} =0$. The associated position probability distribution is
\begin{eqnarray}
w_a({\pmb x}) = \frac{\sum_j |\langle {\pmb x}, j|\omega, a\rangle|^2}{\langle \omega, a|\omega, a\rangle}.
\end{eqnarray}
 Let us denote by $\delta_a$ the localization radius of $w({\pmb x})$, defined by $\int_{|{\pmb x}| > \delta_a} d^3x w_a({\pmb x}) < \epsilon$, for some small number $\epsilon$. If the sampling width of  $\hat{F}_{\pmb x}$ satisfies $\sigma_T >> \delta$, then
\begin{eqnarray}
\hat{F}_{\pmb x} |\omega, a\rangle &=& \int d^3x' \sum_j g({\pmb x} - {\pmb x}') |{\pmb x}', j\rangle \langle {\pmb x}', j| \omega, a\rangle \simeq g({\pmb x}) \int d^3x' \sum_j |{\pmb x}', j\rangle \langle {\pmb x}', j| \omega, a \rangle \nonumber \\
&=& g({\pmb x}) |\omega, a\rangle.
\end{eqnarray}
Hence, $|\omega, a\rangle$ is an approximate eigenvector of $\hat{F}_{\pmb x}$. Using the mean-value theorem, we can show that, for a Gaussian $g$,   the approximation  has an error  of order $ (\delta_a/\sigma_T)^2$.

The kernel (\ref{kernelR}) becomes
\begin{eqnarray}
R^{ab}_{{\pmb x}, t} ({\pmb y}, s; {\pmb y}', s') =    \sqrt{f(t-s)f(t - s')g({\pmb x} - {\pmb y})g({\pmb x} - {\pmb y'})} \langle \omega, b|e^{i \mathfrak{h}(s-s') - i \mathfrak{\pmb p}\cdot ({\pmb y} - {\pmb y'})}| \omega, a\rangle. \label{kernelR2}
\end{eqnarray}

We introduce the spacetime coordinates $X = (t, {\pmb x})$, $ Y= (s, {\pmb y})$, and $ Y' = (s', {\pmb y}')$, and the detector four-momentum  $\hat{\cal P} = ( \hat{\mathfrak{h}}, \hat{\mathfrak{\pmb p}})$. We also define the spacetime sampling function
$\eta(X) = f(t) g({\pmb x})$, normalized to unity $\int d^4X \eta(X) = 1$.

 Then, Eq. (\ref{kernelR2}) becomes
\begin{eqnarray}
R^{ab}_{X}(Y, Y') =   \sqrt{\eta(X-Y)\eta(X-Y')} S^{ab}(Y'-Y) ,
\end{eqnarray}
\noindent
where
\begin{eqnarray}
S^{ab}(Y-Y') \langle \omega, b|e^{-i \hat{\cal P} \cdot Y}|\omega, a\rangle.
\end{eqnarray}

We write Eq. (\ref{probrel})

\begin{eqnarray}
P(X) = \int d^4Y d^4 Y' R^{ab}_{X} (Y,Y') G_{ab}(Y, Y') , \label{covariantprob}
\end{eqnarray}
 in terms of the correlation function
\begin{eqnarray}
G_{ab}(Y, Y') = Tr \left[ \hat{O}_a(Y)\hat{\rho}_0 \hat{O}^{\dagger}_b(Y')\right]. \label{two-point}
\end{eqnarray}

Eq. (\ref{covariantprob}) is the main result of this section. It expresses the probability density $P(X)$ for particle detection as a linear functional of the two-point correlation function (\ref{two-point}).
We emphasize that no particle concepts were introduced in the derivation of Eq. (\ref{covariantprob}). All information about the measured quantum system is contained in the field
 correlation function $G_{ab}(Y, Y')$.

We emphasise that Eq. (\ref{covariantprob}) applies only for measurement events, i.e., it presupposes that a macroscopic measuring apparatus has been included in the quantum description. The variable $X$ is a decoherent  macroscopic pointer variable defined by the apparatus's degrees of freedom. In absence of the apparatus (or analogous decoherence mechanisms \cite{HaZa}) spacetime probabilities like Eq. (\ref{covariantprob}) cannot be defined for general initial states \cite{YaTa}.

\subsection{The detector Hilbert space}

Let ${\cal V} = L^2({\pmb R}^4) \otimes {\pmb C}^n$ be the Hilbert space that contains the test functions $f^a(X)$, that smear the composite operators $\hat{O}_a(X)$,  $a = 1, 2, \ldots, n$. The inner product on ${\cal V}$ is defined by
\begin{eqnarray}
(g,f)_{\cal V} = \int d^4X f^a(X) [g^b(X)]^* m_{ab}, \label{innerproduct}
\end{eqnarray}
for some positive definite matric $m_{ab}$ on ${\pmb C}^n$ that is used to raise and lower the indices $a$ and $b$.

The kernel $R^{ab}_{X}$ defines an operator on ${\cal V}$,
\begin{eqnarray}
(\hat{R}_Xf)^a(Y) = \int d^4Y' R^{ab}_X(Y, Y') m_{bc} f^c(Y').
\end{eqnarray}
By Eq. (\ref{kernelR}), $\hat{R}_X$ is the Gram matrix associated to the vectors  $\sqrt{\eta(X-Y) } e^{i \hat{\cal P} \cdot Y} |\omega, a \rangle$; hence, it is a non-negative operator.   It is also trace-class
\begin{eqnarray}
Tr_{\cal V} \hat{R}_X = \sum_{a,b}  m_{ab}   \langle \omega, a  |\omega, b\rangle < \infty.  \label{trvr}
\end{eqnarray}

Hence, $\hat{R}_X$ is a density matrix on ${\cal V}$, modulo a proportionality constant that can be set to unity by appropriate normalization of the probabilities.

The two-point function $G_{ab}(Y, Y')$
also defines a operator $\hat{G}$ on ${\cal V}$. Being a Gram matrix, $\hat{G}$ is positive, but not necessarily trace-class.

The probability density (\ref{covariantprob}) can be written in the suggestive form
\begin{eqnarray}
P(X) = Tr_{\cal V} \left( \hat{R}_X \hat{G}\right). \label{probmain}
\end{eqnarray}

We emphasize that the Hilbert space ${\cal V}$ where the operators $\hat{R}_X$ and $\hat{G}$ are defined is {\em not} the Hilbert space ${\cal F}$ of field quantum states. ${\cal V}$ is associated to the detector; hence, we call it {\em detector Hilbert space}. Density matrices on ${\cal V}$ define time-extended measurements,  not single-time states. All information about initial state and dynamics of the quantum field is encoded in $\hat{G}$.

\subsection{Probabilities in the Weyl-Wigner picture}

 The Weyl-Wigner (WW) symbol  $\tilde{A} (X, K)$ of any operator $\hat{A}$ on $L^2({\pmb R}^4)$ is a function of the `position' $X^{\mu}$ and its conjugate momentum $K_{\mu}$, defined by
 \begin{eqnarray}
\tilde{ A}(X, K) = \int \frac{d^4\xi}{(2\pi)^4} e^{i K\cdot \xi} (X+\frac{1}{2}\xi|\hat{A}| X - \frac{1}{2}\xi).
 \end{eqnarray}
 A key-property of the Weyl-Wigner transform is that for any two operators $\hat{A}$ and $\hat{B}$ on $L^2({\pmb R}^4)$.
 \begin{eqnarray}
 Tr(\hat{A}\hat{B}) = \int \frac{d^4X d^4K}{(2\pi)^4}  \tilde{A}(X, K) \tilde{B}(X, K).
 \end{eqnarray}

 So far, we have not specified the sampling function $\eta(X)$. We choose Gaussian samplings,
 \begin{eqnarray}
 \eta(X) = \frac{\sqrt{ \det M}}{4\pi^2}\exp(-\frac{1}{2} M_{\mu \nu} X^{\mu}X^{\nu}), \label{Gaussian}
 \end{eqnarray}
 where $M_{\mu \nu}$ is a positive definite matrix on ${\pmb R}^4$. For example, in a Lorentz frame that is defined by a normal unit time-like vector $n = \frac{\partial}{\partial t}$,
 \begin{eqnarray}
 M_{\mu\nu} = \frac{1}{\sigma_T^2} n_{\mu} n_{\nu} + \frac{1}{\sigma_X^2}(n_{\mu} n_{\nu}- \eta_{\mu \nu})
 \end{eqnarray}
 where $\sigma_T$ is the width of the temporal sampling and $\sigma_X$ the width of the position sampling.

  Gaussian sampling functions have two desired properties. First, they introduce no spurious correlation between $X$ and $K$ in the WW symbol of $\hat{R}_X$. Second,  they remain Gaussian sampling functions in all Lorentz frames---the positivity of $M_{\mu \nu}$ is preserved under Lorentz transformations.

For Gaussian sampling functions,  the WW symbol $\tilde{R}_X$ factorizes
  \begin{eqnarray}
  \tilde {R}^{ab}_X(Y, K) =  \eta(Y-X) \tilde{S}^{ab}(K),
  \end{eqnarray}
    where
    \begin{eqnarray}
    \tilde{S}^{ab}(K) = \int \frac{d^4\xi}{(2\pi)^4} e^{- \frac{1}{8}M_{\mu \nu}\xi^{\mu} \xi^{\nu}} e^{ i K\cdot \xi} S^{ab}(\xi). \label{lambdadef}
    \end{eqnarray}

    The function $S^{ab}(Y)$ quantifies correlations between excitations that are localized at different spacetime points. Since the detector is a macroscopic system with a large number  of particles, $S(Y)$  decays for times larger than some characteristic microscopic decoherence time $\tau$ and for distances larger than a microscopic correlation length $\delta$.

   The decoherence condition (\ref{deccond}) is satisfied for samplings, such that
    \begin{eqnarray}
    \sigma_T >> \tau, \hspace{1cm} \sigma_X >> \delta.  \label{sepscales}
     \end{eqnarray}
Assuming Eq. (\ref{sepscales}),       the Gaussian in Eq. (\ref{lambdadef}) can be set to unity. Then, $\tilde{S}^{ab}(K)$ does not depend on the sampling parameters, and it is simply the Fourier transform of  $S^{ab}$,
\begin{eqnarray}
    \tilde{S}^{ab}(K) = \int \frac{d^4\xi}{(2\pi)^4}  e^{ i K\cdot \xi} S(\xi) = \langle \omega, b|\delta^4(\hat{\cal P} - K) |\omega, a\rangle. \label{lambdadef2}
    \end{eqnarray}

Eq. (\ref{probmain})  becomes
\begin{eqnarray}
P(X) = \int d^4Y \eta (Y-X) P_0(Y).
\end{eqnarray}
where
\begin{eqnarray}
P_0(X) =  \int \frac{ d^4K}{(2\pi)^4}\tilde{G}_{ab}(X, K)\tilde{S}^{ab}(K). \label{P0X}
\end{eqnarray}

If $P_0(X)$ is positive, then $P(X)$ is  the convolution of an underlying  probability distribution $P_0(X)$ due to finite-width sampling. However, unlike $P(X)$, $P_0(X)$ is not guaranteed to be positive  for any   $G_{ab}$ and any  $\tilde{S}^{ab}$.

We expect that in the regime where Eq. (\ref{lambdadef2})  holds, $P_0(X)$ is  positive. A heuristic argument is the following.
 Eq. (\ref{lambdadef2}) implies that $P_0(X)$   is independent of the sampling parameters. But then $\sigma_X$ and $\sigma_T$ appear in $P(X)$ only through  $\eta$, and they can be varied arbitrarily. For sufficiently small values, $P_0(X)$ comes arbitrarily close to $P(X)$, and is therefore positive.    This expectation is physically intuitive. It  is  common in studies of emergent classicality \cite{Omn1, GeHa, An95}, where  a  separation of scales like Eq. (\ref{sepscales})   implies the definition of sampling-independent
  probability distributions.

We will refer to detectors characterized by positive $P_0(X)$ for a large class of physically relevant states as {\em regular detectors}, meaning that they can be assigned probabilities that are independent of the sampling.

Equation (\ref{P0X}) applies to any QFT (free or interacting), and  for any field-apparatus coupling. In what follows, we will consider measurements of relativistic spinless particles.

\section{Measurements on a scalar field }

 We apply the measurement theory of Sec. 2, to a QFT containing only a single scalar field
 $\hat{\phi}(X)$, describing particles of mass $m$.    Possible choices for the composite operators $\hat{O}_a(X)$ that describe the field-apparatus coupling are  the following.

\begin{enumerate}[(i)]
\item $\hat{O}(X) = \hat{\phi}(X)$. This coupling characterizes Unruh-DeWitt detectors \cite{Dewitt}. It is the scalar-field  analogue of the electromagnetic (EM) dipole coupling. It describes particle detection by absorption.

\item $\hat{O}(X) = \hat{\phi}^{(+)}(X)$. The detector couples to the positive frequency part  $\hat{\phi}^{(+)}(X)$ of the quantum field. This is an analogue of the Glauber EM field-detector coupling in quantum optics \cite{Glauber}. A drawback of this coupling is that the splitting of the field into positive and negative frequency parts is a non-local operation; hence, it could lead to causality violation in set-ups that involve multiple measurements.

\item    $\hat{O}(X) = :\hat{\phi}^2(X):$. This coupling describes particle detection by scattering through a scalar interaction. The double dots denote normal ordering.

\item $\hat{O}_{\mu}(X) = :\hat{\phi}(X)\partial_{\mu} \hat{\phi}(X):$. This coupling describes particle detection by scattering through a vector interaction.
\end{enumerate}

In what follows, we  consider   {\em scalar} composite operators $\hat{O}(X)$, i.e., cases (i)---(iii) above. We drop the indices $a, b$ that appear in the detector kernel $R_X(Y, Y')$,  the two-point function $G(X, X')$, and the correlation function $S(Y)$.   The probability formula, Eq. (\ref{P0X}) becomes
 \begin{eqnarray}
P_0(X) =  \int \frac{ d^4K}{(2\pi)^4}\tilde{G}(X, K)\tilde{S}(K), \label{P0X2}
\end{eqnarray}

The detector Hilbert space is $L^2({\pmb R}^4)$.  The matrix $m_{ab}$ of Eq. (\ref{innerproduct}) is unity, and there is a single vector $|\omega\rangle = \hat{J}(0)|0\rangle$.
 Eq. (\ref{trvr}) becomes $Tr_{\cal V} \hat{R}_X = \langle \omega|\omega\rangle$. Hence,
 by normalizing $|\omega\rangle$ to unity, $Tr_{\cal V} \hat{R}_X  = 1$. With this normalization, $S(0) = 1$.

 A  quantum scalar field that satisfies a linear wave equation can be analysed in terms of creation and annihilation operators
 \begin{eqnarray}
 \hat{\phi}(X) = \sum_a \left[ \hat{a}_a \chi_a(X) + \hat{a}^{\dagger}_a \chi_a^*(X)\right], \label{fieldphi}
 \end{eqnarray}
 where $\chi_a(X)$ are positive-frequency mode solutions to the wave equation, $a$ an index that labels the modes, and the creation and annihilation operators satisfy the canonical commutation relations,
   \begin{eqnarray}
   [\hat{a}_a, \hat{a}_b] =    [\hat{a}^{\dagger}_a, \hat{a}^{\dagger}_b] = 0, \hspace{0.1cm}  [\hat{a}_a, \hat{a}^{\dagger}_b] = \delta_{ab}.
   \end{eqnarray}
Eq. (\ref{fieldphi}) applies also in presence of external time-dependent fields and non-trivial boundary conditions.

For a Glauber-type  coupling, $\hat{O}(X) = \hat{\phi}^{(+)}(X)$, the two-point function (\ref{two-point}) is
\begin{eqnarray}
G_0(X, X') = \sum_{ab} \rho_{ab} \chi_a(X) \chi_b^*(X'), \label{GXX}
\end{eqnarray}
where $\rho_{ab} = Tr\left( \hat{a}_a\hat{\rho}_0 \hat{a}_b^{\dagger}\right)$ is the one-particle reduced density matrix. We assumed an initial state with a definite number of particles.

The one-particle density matrix can be determined in terms of external operations on the quantum field vacuum $|\Omega\rangle$. For example, assume that the system is initially ($t \rightarrow -\infty$) in the vacuum state and it is acted upon by an classical external source $J(X)$ with support at times $t < 0$. Then the quantum field state at $t = 0$ is $\exp[ -i \int d^4X J(X) \hat{\phi}(X)]|\Omega\rangle$, and the one-particle reduced density matrix is pure, $\rho_{ab} = \psi_a \psi^*_b$, with
\begin{eqnarray}
\psi_a = \int d^4X J(X) \chi_a(X).
\end{eqnarray}

For an Unruh-deWitt type coupling,  $\hat{O}(X) = \hat{\phi}(X)$, the two-point function  (\ref{two-point}) is
\begin{eqnarray}
G(X,X') = G_0(X, X') + G_0(X', X) + \sum_a \chi_a(X) \chi_a^*(X'), \label{Wight}
\end{eqnarray}
where $G_0(X, X')$ is given by Eq. (\ref{GXX}).  The last term is state-independent and corresponds to particle creation by the vacuum. In absence  of strong time-dependent external fields, it is a small background noise term that can be ignored. Given the fact that the probabilities are defined up to a normalization constant, the term $G_0(X,X') + G_0(X', X)$ leads to the same probabilities with $G_0(X, X')$. Hence,  we can use the two-point function (\ref{GXX}) for both Glauber and Unruh-DeWitt couplings.

 For a free field in Minkowski spacetime, the index $a$ corresponds to particle three-momentum ${\pmb p}$, the summation $\sum_a$ to $\int   d^3p  $ for $\epsilon_{\pmb p} = \sqrt{m^2 + {\pmb p}^2}$, and the mode functions are
 \begin{eqnarray}
 \chi_{\pmb p} = \frac{1}{(2\pi)^{3/2}\sqrt{2\epsilon_{\pmb p}}} e^{-i P\cdot X}, \label{mode3d}
 \end{eqnarray}
where $P^{\mu} = (\epsilon_{\pmb p}, {\pmb p})$. Hilbert space vectors $\psi({\pmb p})$ are normalized as $\int    d^3p  |\psi({\pmb p})|^2 = 1$.

For a pure state $\psi({\pmb p})$, the probability density (\ref{P0X}) becomes
\begin{eqnarray}
P_0(X) = \int d^4 \xi \psi_{Ph}(X - \frac{1}{2} \xi) \psi_{Ph}^*(X + \frac{1}{2} \xi) S(\xi), \label{pnw}
\end{eqnarray}
where
\begin{eqnarray}
\psi_{Ph}(X) := \int \frac{d^3p}{(2\pi)^{3/2}} \frac{\psi({\pmb p}) }{\sqrt{2\epsilon_{\pmb p}}} e^{-i P\cdot X},
\end{eqnarray}
is the Phillips wave function \cite{Phil}.

For the scattering coupling (iii) and for a single-particle initial state, the two-point function
  (\ref{two-point}) is
  \begin{eqnarray}
  G(X,X') = G_0(X, X')  \left[\sum_a \chi_a(X) \chi_a^*(X')\right].
  \end{eqnarray}

 For a free field, substitution into Eq. (\ref{P0X2}) yields
 \begin{eqnarray}
 P_0(X) =  \int \frac{ d^4K}{(2\pi)^4}\tilde{G}_0(X, K)\tilde{S}_1(K),
 \end{eqnarray}
 where
 \begin{eqnarray}
 \tilde{S}_1(K) = \int \frac{d^3q}{(2\pi)^3 (2 \epsilon_{\pmb q})} \tilde{S}(Q+K),
\end{eqnarray}
where $Q^{\mu} = (\epsilon_{\pmb q}, {\pmb q})$.

Hence, we can use the two point function (\ref{GXX}) also for a coupling of type (iii), modulo a redefinition of the  function $\tilde{S}$.


\section{Time-of-arrival measurements}
 In this section, we  construct  probability distributions for time-of-arrival measurements on spinless particles.

\subsection{Localization operator}
In a  time-of-arrival measurement, a detector is placed at a macroscopic distance $x$ from the particle source. If $x$ is much larger than the size of the detector, only particles with momenta along the axis that connects the source to the detector are recorded. Hence, the problem is  reduced to two spacetime dimensions. We set the spacetime coordinate $X = (t, x)$, the mode functions are labeled by a single momentum coordinate, the summation over $a$ in (\ref{GXX}) becomes  $\int dp$, and the on-shell four-momentum is $P^{\mu} = (\epsilon_p, p)$ with $\epsilon_p = \sqrt{p^2+m^2}$. We also express the off-shell momentum $K^{\mu} = (E, K)$,  and write the function $\tilde{S}(K^{\mu})$ as  $\tilde{S}(K, E)$;  $E$ and $K$ are independent parameters. Then, Eq. (\ref{P0X}) becomes
\begin{eqnarray}
P_0(t, x) = \int \frac{dpdp'}{2\pi } \frac{\rho(p,p')}{2\sqrt{\epsilon_p \epsilon_{p'}}} \; \tilde{S}\left( \frac{p+p'}{2}, \frac{\epsilon_p + \epsilon_{p'}}{2}\right) e^{i(p-p')x - i (\epsilon_p - \epsilon_{p'})t}, \label{ptx}
\end{eqnarray}
modulo an overall multiplicative constant.

 We normalize Eq. (\ref{ptx}) following a prescription, common to non-relativistic treatments of the time of arrival \cite{Kij}. We fix the value of $x > 0$, so that  the location of the detector is a parameter and not a random variable.  Then, we require that
 \begin{eqnarray}
 \int_{-\infty}^{\infty} dt P_0(t, x) = 1, \label{normal}
 \end{eqnarray}
 for all states with support only on positive momenta.  The integration is extended to the full real axis for time, because $P_0(t, x)$ for $t <0$ is negligibly small for any wave function with strictly positive momentum.

 We evaluate the time-integrated probability density,
 \begin{eqnarray}
 p_{tot} = \int_{-\infty}^{\infty} dt P(t,x) = \int_0^{\infty} dp \rho(p,p) \frac{\tilde{S}(p, \epsilon_p)}{2p}. \label{normalization}
 \end{eqnarray}
 Hence,
 \begin{eqnarray}
  \frac{\tilde{S}(p, \epsilon_p)}{2p} = \alpha(p), \label{alphap}
 \end{eqnarray}
 where $\alpha(p)$ is the {\em absorption coefficient} of the detector, namely, the fraction of particles at momentum $p$ that are absorbed when crossing the detector\footnote{For an
 elementary detecting element of length $d$, the absorption coefficient is the product of the linear attenuation coefficient of the material with $d$. It is therefore a directly measurable quantity.
 }.  The normalization condition  (\ref{normal})  is implemented by dividing $P_0(t, x)$ with $p_{tot}$.
 We also redefine the initial state by
 \begin{eqnarray}
 \rho(p, p') \rightarrow \tilde{\rho}(p, p') = \frac{ \sqrt{\alpha(p) \alpha(p')} \rho(p,p')}{\int_0^{\infty} dp \rho(p,p) \alpha(p)}.
\end{eqnarray}
Then,
 \begin{eqnarray}
P_0(t, x) = \int \frac{dpdp'}{2\pi } \tilde{\rho}(p,p')  \sqrt{v_p v_{p'}} L(p,p') e^{i(p-p')x - i (\epsilon_p - \epsilon_{p'})t}, \label{ptxb}
\end{eqnarray}
where $v_p = p/\epsilon_p$ is the particle velocity.  $L(p, p')$ are the matrix elements $\langle p|\hat{L}|p'\rangle$ of the {\em  localization operator} $\hat{L}$, defined by

\begin{eqnarray}
 \langle p|\hat{L}|p'\rangle  := \frac{\tilde{S}\left( \frac{p+p'}{2}, \frac{\epsilon_p + \epsilon_{p'}}{2}\right)}{\sqrt{\tilde{S}(p, \epsilon_p) \tilde{S}(p', \epsilon_{p'})}}.  \label{lpp}
\end{eqnarray}
By definition, $\langle p|\hat{L}|\hat{p'}\rangle \geq 0 $, and
 \begin{eqnarray}
 L(p, p) = 1. \label{norm2}
\end{eqnarray}

For a pure state initial state $|\psi\rangle$, Eq. (\ref{ptx}) becomes
\begin{eqnarray}
P_0(t, x) = \langle \psi|\hat{U}^{\dagger}(t, x) \sqrt{|\hat{v}|}\hat{L}\sqrt{|\hat{v}|}\hat{U}(t, x)|\psi\rangle, \label{ptxc}
\end{eqnarray}
where $\hat{U}(t, x) = \hat{U}(X)$ is the  spacetime-translation operator
\begin{eqnarray}
\hat{U}(t, x) = e^{i  \hat{p} x - i \hat{H} t} = e^{ -i X\cdot \hat{P}},
\end{eqnarray}
and $\hat{v} = \hat{p}\hat{H}^{-1}$ is the velocity operator.

In regular detectors, $P_0(t, x)$ is positive for all initial states. Εq. (\ref{ptxc}) implies that $\hat{L}$ is a positive operator. Then, the Cauchy-Schwarz inequality applies,
\begin{eqnarray}
\langle p|\hat{L}|p'\rangle \leq \sqrt{\langle p|\hat{L}|p\rangle \langle p'|\hat{L}|p'\rangle} =  1. \label{CSa}
\end{eqnarray}

\noindent  Eq. (\ref{CSa}) is always satisfied if $\ln \tilde{S}(p, \epsilon_p)$ is  a  convex  function of $p$.

 The   operator $\hat{L}$ determines the   localization of the detection event.   To see this, consider the Weyl-Wigner transform $\tilde{L}(x, p)$ of  $\hat{L}$,
 \begin{eqnarray}
\tilde{L}(x, p) = \int \frac{d \xi}{2 \pi} \langle p+\frac{\xi}{2}|\hat{L}|p-\frac{\xi}{2}\rangle e^{-i\xi x}
 \end{eqnarray}

  By Bochner's theorem \cite{RS2}, the positivity of the matrix elements $\langle p|\hat{L}|p'\rangle$ implies that $\tilde{L}$ is non-negative. Furthermore,
 \begin{eqnarray}
\int dx \tilde{L}(x, p)  = \langle p|\hat{L}|p\rangle = 1.
 \end{eqnarray}
 Hence, $\tilde{L}(x, p)$ is a probability density with respect to $x$ in which $p$ appears as a parameter, and not a random variable. To highlight  this key point, we will write $\tilde{L}(x, p)$ as $u_p(x)$. The probability density $u_p(x)$ describes the irreducible spread of the measurement record due to the physics of the detector. If $u_p(x)$ is $p$-independent, i.e., the spread is the same at all momenta, then $L(p,p')$ is a function of $p-p'$.

   The Fourier transform $\tilde{u}_p(\xi)$ equals $\langle p - \frac{\xi}{2}|\hat{L}|p + \frac{\xi}{2}\rangle$. It is straightforward to show that $\tilde{u}'_p(0) = 0$, which means that the average of $x$ vanishes.
 The  associated variance $\sigma^2(p):= - \tilde{u}_p''(0)$ quantifies the width of the detection record,
\begin{eqnarray}
\sigma^2 (p) =  \frac{1}{4} \frac{d^2 \ln \tilde{S}(p, \epsilon_p)}{d p^2}  - \frac{m^2}{4\epsilon_p^3} (\partial_E\ln \tilde{S})(p, \epsilon_p) =
\frac{1}{4} (\ln \alpha)''(p) - \frac{1}{4 p^2} - \frac{m^2}{4\epsilon_p^3} (\partial_E\ln \tilde{S})(p, \epsilon_p).
\label{detectorsigma}
\end{eqnarray}

The probability density (\ref{ptxb}) is defined only for initial states $\hat{\rho}$ with support on positive momenta. It can straightforwardly be extended to all initial states, by inserting a multiplicative term $\theta(pp')$ into the integral (\ref{ptxb}), and extending the integration over the full range of $p$ and $p'$ \cite{Kij}. The resulting probability density is normalized to unity. This prescription drops the contributions from off-diagonal elements of the density matrix $\rho(p, p')$ with $pp' < 0$. The contribution of such terms to the total probability is negligible for  macroscopically large source-detector distance $x$.

 \subsection{Maximal localization}
 Maximal localization is achieved when  Eq. (\ref{CSa}) is saturated, i.e.,  for  $\langle p|\hat{L}|p'\rangle =  1$. This means that $\hat{L} = \delta(\hat{x})$, where $\hat{x}$ is the Newton-Wigner position operator $i \frac{\partial}{\partial p}$. Equivalently,  $u_p(x) = \delta(x)$.

 The probability density for maximal localization is
  \begin{eqnarray}
  P_{m.l.}(t,x) = \langle \psi|\hat{U}^{\dagger}(t, x)\sqrt{|\hat{v}|}\delta(\hat{x})\sqrt{|\hat{v}|}  \hat{U}(t, x)|\psi\rangle, \label{ptxmax}
  \end{eqnarray}
or
\begin{eqnarray}
 P_{m.l.}(t, x) = \left| \int \frac{dp}{2\pi } \psi(p) \sqrt{\frac{|p|}{\epsilon_p}} e^{ipx - i \epsilon_p t} \right|^2. \label{POVM3}
\end{eqnarray}

Eq.  (\ref{POVM3}) was identified in Ref. \cite{AnSav12} using the QTP method. The same expression had previously been found by Le\'on \cite{Leon}, in terms of the eigenvectors of a non-self-adjoint time-of-arrival operator. In the non-relativistic limit, Eq. (\ref{POVM3}) reduces to Kijowski's probability distribution \cite{Kij}.

The condition $\langle p|\hat{L}|p'\rangle =  1$ that leads to Eq.
 (\ref{ptxmax}) implies that $\ln \tilde{S}$ is a linear function, i.e.,
 \begin{eqnarray}
 \tilde{S}(E, K) = C e^{- \tau E - \delta K},   \;\; E, K > 0.
 \end{eqnarray}
where $C$ is a positive constant, $\tau$ is the decoherence time and $\delta$ the localization length of the record. This implies that
\begin{eqnarray}
S(t, x) = \frac{1}{(1 + \frac{it}{\tau})(1 - \frac{i x}{\delta})},
\end{eqnarray}
where the constant $C = \tau \delta $ was determined by the requirement that $S(0, 0)= 1$.

By Eq. (\ref{alphap}), the absorption coefficient for maximal localization is
\begin{eqnarray}
\alpha(p) = \frac{\tau \delta e^{- \tau \epsilon_p - \delta p}}{2 p}.
\end{eqnarray}

 Eq. (\ref{ptxmax}) can also be written as
\begin{eqnarray}
 P_{m.l.}(t, x) = |\langle 0_x|\sqrt{\hat{v}_p} \hat{U}(x, t)|\psi\rangle|^2. \label{ptxmax2}
\end{eqnarray}
where $| 0_x \rangle$ is the generalized eigenvector of $\hat{x}$ with zero eigenvalue: $\langle p|0_x\rangle = (2\pi)^{-1/2}$.

 The probability density $P_{m.l.}(t, x)$ is the only one that can be factorized as the modulus square of an amplitude. Factorization requires a localization operator of the form $L(p, p') = c(p) c(p')$, for real-valued functions $c(p)$. The condition $L(p, p) = 1$ implies that $c(p)^2 = 1$, hence $L(p, p') =1$.

The probability density (\ref{ptxmax}) cannot be expressed as  a local functional of any wave-function $\psi(t, x)$. However, for almost monochromatic initial states, it approximates a probability current
$- \mbox{Im}  \psi^*_{Ph}(t,x) \partial_x \psi_{Ph}(t, x)$ with respect to the Phillips wave function
\begin{eqnarray}
\psi_{Ph}(t, x) := \int \frac{dp}{\sqrt{2\pi} \epsilon_p} \psi(p)  e^{ipx-i \epsilon_pt}.
\end{eqnarray}

 \subsection{Other detector types}
The localization operator $\hat{L}$ is determined by a function $\tilde{S}(K, E)$   of two independent variables. Of particular interest are the cases  that  $\tilde{S}$ depends non-trivially on only one  variable. Then, $\hat{L}$ is uniquely determined by the absorption coefficient.

\medskip

\noindent {\em 1. Fully decoherent detectors.} Assume that the spatial localization scale $\delta$ is much larger than the decoherence time  $\tau$, so that we can take $\tau \rightarrow 0$. Then, $\tilde{S}$ is  a function only of $K$.
 By Eq. (\ref{alphap}), $\tilde{S}(K) = 2K \alpha(K)$, hence,
\begin{eqnarray}
\langle p|\hat{L}|p'\rangle = \frac{p+p'}{2\sqrt{pp'}} \frac{\alpha(\frac{1}{2}(p+p'))}{\sqrt{\alpha(p)\alpha(p')}}. \label{locspc}
\end{eqnarray}
The detection width  is
\begin{eqnarray}
\sigma^2 (p) =
\frac{1}{4} (\ln \alpha)''(p) - \frac{1}{4 p^2}.
\end{eqnarray}

In the non-relativistic limit and for $\alpha = 1$, the localization operator (\ref{locspc}) corresponds to the time-of-arrival probability distribution generated by the probability current  and by a non-self adjoint time operator \cite{DeMu97, mals, BEMS, DEHM}. However, for $\alpha = 1$, $\hat{L}$ is not a positive operator. Not only is  $\sigma^2(p)$ negative but it can be made arbitrarily large in absolute value for states with momentum support near zero.

\medskip

\noindent {\em 2. Coherent detectors.} In the regime  $\delta << \tau$,   $\tilde{S}$ is a function only of $E$. By Eq. (\ref{alphap}), $\tilde{S}(E) = 2\sqrt{E^2-m^2}\alpha(\sqrt{E^2-m^2})$, hence,
\begin{eqnarray}
\langle p|\hat{L}|p'\rangle = \frac{\sqrt{\frac{1}{4} (\epsilon_p+\epsilon_{p'})^2-m^2}}{\sqrt{pp'}} \frac{\alpha( \sqrt{\frac{1}{4} (\epsilon_p+\epsilon_{p'})^2-m^2})}{\sqrt{\alpha(p)\alpha(p')}}. \label{locspcc}
\end{eqnarray}

The detection width  is
\begin{eqnarray}
\sigma^2 (p) =
\frac{1}{4} (\ln \alpha)''(p) - \frac{m^2}{4 \epsilon_p^3} (\ln \alpha)'(p) - \frac{1}{4 \epsilon_p^2}.
\end{eqnarray}

 In the non-relativistic limit and for $\alpha = 1$, the localization operator (\ref{locspcc}) corresponds to the time of arrival probability distribution derived in Ref. \cite{AnSav17}. Again, for $\alpha = 1$, the localization operator is non-positive. However, its negative values are negligible in the non-relativistic limit:  $\sigma^2(p) $  is always greater than $-(4m)^{-2}$. For example, the negative contributions of $\sigma^2(p) $   to the time-of-arrival variance  (\ref{uncert}) can be ignored.

\medskip

\noindent {\em 3. Covariant detectors.} If the initial state  $|0\rangle$ of the detector is Lorentz invariant (e.g., it is a QFT vacuum),  and the  current operator $\hat{J}(X) $ is a Lorentz scalar, then $S$ is a Lorentz scalar. Its Fourier transform
 $\tilde{S}$
is a function only of $z = E^2 - K^2$. We therefore write $\tilde{S}$ as $\Sigma(z)$. Eq. (\ref{alphap}) implies that $\alpha(p) = \Sigma(m^2)/(2p)$, hence,
\begin{eqnarray}
\langle p|\hat{L}|p'\rangle = \frac{\Sigma\left[\frac{1}{2}(m^2+P\cdot P')\right] }{\Sigma(m^2)}.
\end{eqnarray}
The detection width is
\begin{eqnarray}
\sigma^2(p) = (\ln \Sigma)'(m^2)\frac{m^2}{\epsilon_p^2}.
\end{eqnarray}

\medskip
\noindent {\em 4. Ideal detectors.} Any momentum-dependent probability distribution $u_p(x)$ defines a localization operator, and, hence, a POVM for the time of arrival. In an ideal detector, the localization operator does not depend on any parameter that characterizes the detector, such as the correlation length $\delta$ or the decoherence time $\tau$. This is the case, for example, if  $u_p(x) = p f(px)$, where $f(s)$ is a probability distribution with respect to the dimensionless variable $s$. Then, $L(p, p')$ is a function solely of the variable $\frac{p-p'}{p+p'}$. A class of such POVMs in the non-relativistic regime has been studied in Ref. \cite{AnSav17}.

\subsection{Poincar\'e covariance and associated position operators}
Next, we examine how the time-of-arrival probability densities transform under the action of the Poincar\'e group.

The probability density $P_0(X)$  Eq. (\ref{ptx}),  transforms covariantly under spacetime translations. The transformation $|\psi\rangle \rightarrow e^{-i\hat{p}a + i \hat{H}b} |\psi\rangle $ induces a transformation $P_0(t, x) \rightarrow P_0(t - a, x - b)$.

A  Lorentz transformation $\hat{U}(\Lambda)$  transforms a  state $\psi(p)$ to $\sqrt{\frac{\epsilon_{\Lambda^{-1} p}}{\epsilon_p}}\psi(\Lambda^{-1}p)$. Then, $P_0(X)$  is a Lorentz scalar, $P_0(X) \rightarrow P_0(\Lambda X)$, only if  $\tilde{S}$ is Lorentz invariant. As shown in Sec.  4.3, this property characterizes only a special class of detectors. It is also not a necessary condition, an actual particle detector defines a specific Lorentz frame (the laboratory frame).

However, even if $P_0(X)$ is a spacetime scalar, the  probability density (\ref{ptxb}) for the time-of-arrival is not. In deriving Eq. (\ref{ptxb}), we restricted to a statistical sub-ensemble of particles detected at   $x$, and we transformed the state accordingly. This restriction broke the Lorentz symmetry. Indeed, in Eq. (\ref{ptxb}), the time $t$ is a random variable and $x$ is a parameter, so, by definition, $P_0(t, x)$ cannot be a spacetime scalar.

This remark also explains the non-covariance of relativistic position operators. A position operator describes a subensemble of particles detected at any $x$, but at a fixed moment of time $t$. The normalization condition relevant to this subensemble is  $\int dx P(t, x) = 1$.
By Eq. (\ref{ptx}),  $\int dx P_0(t, x) = \int dp \rho(p, p) v_p \alpha(p)$, where we used Eq. (\ref{alphap}). We redefine the initial state
\begin{eqnarray}
 \rho(p, p') \rightarrow \tilde{\rho}(p, p') = \frac{ \sqrt{v_pv_{p'}\alpha(p) \alpha(p')} \rho(p,p')}{\int_0^{\infty} dp \rho(p,p) v_p\alpha(p)}.
  \end{eqnarray}
Then, Eq. (\ref{ptx}) becomes
\begin{eqnarray}
P_0(t, x) = \int \frac{dpdp'}{2\pi } \tilde{\rho}(p,p')   L(p,p') e^{i(p-p')x - i (\epsilon_p - \epsilon_{p'})t}. \label{ptxb3}
\end{eqnarray}
For  a pure initial state $|\psi\rangle$,
\begin{eqnarray}
P_0(t, x) = \langle \psi_t| e^{-i\hat{p}x}\hat{L} e^{i \hat{p}x}|\psi_t\rangle, \label{pxtr}
\end{eqnarray}
where $|\psi_t\rangle  = e^{-i \hat{H}t} |\psi\rangle$. Hence, the translated localization operator
\begin{eqnarray}
\hat{P}_x = e^{-i\hat{p}x}\hat{L} e^{i \hat{p}x}
\end{eqnarray}
 defines a position POVM. For maximal localization, $\hat{P}_x = \delta(\hat{x}-x)$ corresponds to  projective position measurements, and Eq. (\ref{pxtr}) gives the Newton-Wigner probability distribution.

Hence,  there is  a duality between position and time-of-arrival measurements. Each detector defines  particular localization operator $\hat{L}$. When restricting to a sub-ensemble of particles recorded at fixed location $x$, the localization operator defines a time-of-arrival POVM by Eq. (\ref{ptxc}). When   restricting to a sub-ensemble of particles recorded at fixed time  $t$, the localization operator defines a position  POVM by Eq. (\ref{pxtr}). The two sub-ensembles are distinct, and they require  {\em different transformations} of the initial state in order to define normalized probabilities. This is the reason why neither the time-of-arrival nor the position probabilities transform covariantly under the Lorentz group, even when the probability density $P_0(X)$ of Eq. (\ref{ptx}) does.

\section{Uncertainty relations and particle localization}
In this section, we show that the variance of the time of arrival defines a measure for the localizability of relativistic particles.
\subsection{Moment generating function}

First, we evaluate the moment-generating function $Z(\mu, x) = \int dt P(t, x) e^{i \mu t}$
associated to Eq.  (\ref{ptx}),
\begin{eqnarray}
Z(\mu, x) = \int dpdp' \rho(p,p') \sqrt{v_p v_{p'}} \langle p|\hat{L} |p'\rangle e^{i(p-p')x} \delta(\epsilon_p - \epsilon_{p'} - \mu).    \label{zmx}
\end{eqnarray}
We solve the equation $\epsilon_p - \epsilon_{p'}  = \mu$, by expressing  $\xi := p - p'$ as a function of $\bar{p} = \frac{1}{2}(p + p')$,
\begin{eqnarray}
\xi = \frac{\mu}{v_{\bar{p}}}\left[ 1 +  g(\mu, \bar{p})\right],
\end{eqnarray}
where   $g(\mu, \bar{p}) =  \sum_{n=1}^{\infty} c_n(\bar{p}) \left(\frac{\mu}{\bar{p}}\right)^{2n}$ is a series in  $\mu^2$. The explicit form of  $g(\mu, \bar{p})$ will  not be needed in this paper, only the fact that $ \left( \partial g/\partial \mu\right)_{\mu=0} = 0$.
  Eq. (\ref{zmx}) becomes
\begin{eqnarray}
Z(\mu, x) = \int dpdp' \rho(p,p') \frac{2\sqrt{v_p v_{p'}}}{v_p+v_{p'}} \langle p|\hat{L} |p'\rangle e^{i(p-p')x}  \delta\left(p-p' - \frac{\mu}{v_{\bar{p}}}\left[ 1 +  g(\mu, \bar{p})\right]\right).   \label{zmx3}
\end{eqnarray}
We express $\rho(p, p')$ in terms of its associated Wigner function $W(\bar{x}, \bar{p})$. For this calculation, we expand the term $\frac{2\sqrt{v_p v_{p'}}}{v_p+v_{p'}}$ as a power series in $\xi^2$,
\begin{eqnarray}
\frac{2\sqrt{v_p v_{p'}}}{v_p+v_{p'}} = 1 - \sum_{n=1}^{\infty} d_n(\bar{p})\xi^{2n}.
\end{eqnarray}
In what follows, we will need only  the first term in the series,
\begin{eqnarray}
d_1(\bar{p}) =  \frac{m^4}{8\epsilon_{\bar{p}}^4 \bar{p}^2}.
\end{eqnarray}
Eq. (\ref{zmx3}) becomes
\begin{eqnarray}
Z(\mu, x) = \int d\bar{x} d \bar{p}  W(\bar{x}, \bar{p}) e^{iT_c\mu \left[ 1 +  g(\mu, \bar{p})\right]} \left[ 1 - \sum_{n=1}^{\infty} d_n(\bar{p})\left(\frac{\mu}{v_{\bar{p}}}\right)^{2n}\left[ 1 +  g(\mu, \bar{p})\right]^{2n}\right] \nonumber \\ \times \tilde{u}_{\bar{p}} \left( \frac{\mu}{v_{\bar{p}}}\left[ 1 +  g(\mu, \bar{p})\right]\right), \label{zmx4}
\end{eqnarray}
where
\begin{eqnarray}
T_c (\bar{x}, \bar{p}) = \frac{x - \bar{x}}{v_{\bar{p}}}
\end{eqnarray}
is the classical time-of-arrival observable.

All moments of the time of arrival distribution can be obtained by successive differentiations of the moment generating functional (\ref{zmx4}).

\subsection{Uncertainty relation}

We evaluate the expectation value of the arrival time $\langle t\rangle = - i \left(\frac{\partial \ln Z}{\partial \mu}\right)_{\mu=0}$,
\begin{eqnarray}
\langle t\rangle = \langle T_c\rangle,
\end{eqnarray}
where the expectation of any classical observable
$F$ is $\langle F \rangle =  \int d\bar{x} d \bar{p} W(\bar{x}, \bar{p})F (\bar{x}, \bar{p}) $. Hence, the quantum expectation value agrees with the classical one.

The expectation value $\langle T_c\rangle$ is of the form $Tr (\hat{\rho}\hat{T}_c)$, where
\begin{eqnarray}
\hat{T}_c = \frac{1}{2}\left[ (x - \hat{x})\hat{v}^{-1} + \hat{v}^{-1} (x - \hat{x})\right].
\end{eqnarray}
Like its non-relativistic counterpart, the operator $\hat{T}_c$ is symmetric but not self-adjoint \cite{Kij14}. However, its domain $D_{\hat{T}_c}$ includes all vectors with {\em strictly positive} momentum content, i.e.,  functions $\psi(p)$ with positive momentum that vanish faster than any power of $p$ as $p\rightarrow 0$.

Next, we evaluate the variance $(\Delta t)^2 = - \left(\frac{\partial^2 \ln Z}{\partial \mu^2}\right)_{\mu=0}$ to the time of arrival
\begin{eqnarray}
(\Delta t)^2 &=& (\Delta T_c)^2 + 2\langle d_1(p)v_p^{-2} \rangle - \langle \tilde{u}''_p(0)v_p^{-2}\rangle \nonumber \\
&=& (\Delta T_c)^2  + \frac{m^4}{4} \langle \frac{1}{\epsilon_p^2p^4}\rangle + \langle \frac{\sigma^2(p)}{v_p^2}\rangle, \label{uncert}
\end{eqnarray}
where $\sigma^2(p)$ is given by Eq. (\ref{detectorsigma}).

The first term in the right-hand-side of Eq. (\ref{uncert}) is the variance of $\hat{T}_c$. For any  $|\psi\rangle \in D_{\hat{T}_c}$, $[\hat{H}, \hat{T}_c]|\psi\rangle = i |\psi\rangle$, and therefore,     $\Delta T_c > \frac{1}{2 \Delta H}$. Since $\sigma^2(p) \geq 0 $, we obtain the inequality

\begin{eqnarray}
(\Delta t)^2 \geq \frac{1}{4 (\Delta H)^2} + \frac{m^4}{4} \langle \frac{1}{\epsilon_p^2p^4}\rangle \label{ineqq}
\end{eqnarray}
that applies for all   detectors and for all initial states with strictly positive momentum. All terms that appear in the r.h.s. of Eq. (\ref{ineqq}) are determined solely by the momentum probability distribution associated to the quantum state.

Eq. (\ref{ineqq}) does not define a speed limit of time evolution \cite{timenerv}, as in the common
  interpretation of the time-energy uncertainty relation.  Here, $\Delta t$ is the deviation of the time-of-arrival, a quantity that is straightforwardly defined in a time-of-arrival experiment.

  The r.h.s. of Eq. (\ref{ineqq}) never vanishes and it is finite for all states with positive momentum content. Furthermore, it does not depend on the properties or the location of the detector.
   Therefore, it defines an operational measure for the maximal localization of a quantum state.

  The second term in the r.h.s. of Eq. (\ref{ineqq})  is crucial for this interpretation, because it never vanishes. The first term dominates for almost monochromatic states with $\langle H \rangle >> \Delta H$.   However, for any energy probability distribution that drops slower than $E^{-3}$, $\Delta H = \infty$,    and localization is solely determined by the second term. For example, assume that
  the energy $E \geq m$ follows a
  L\'evy distribution
  \begin{eqnarray}
  P(E) = \sqrt{\frac{c_E}{2\pi}} \frac{e^{- \frac{c_E}{2(E-m)}}}{(E-m)^{3/2}},
  \end{eqnarray}
  with scale parameter $c_E$. Then, $\Delta H  = \infty$. However, $\langle \epsilon_p^{-2}p^{-4} \rangle$ is non-zero.  In the non-relativistic limit ($c_E << m$),   $\Delta t > \frac{\sqrt{3}}{2c_E}$, and in the ultra-relativistic limit ($c_E >> m$), $\Delta t > 51.0 \frac{m^2}{c_E^3}$. Localization is determined by the scale parameter $c_E$.

\subsection{Implications}

 In the non-relativistic limit, the second term in the r.h.s. of Eq. (\ref{ineqq})  becomes $\frac{1}{16}\langle \hat{E}_k^{-2}\rangle$, where $\hat{E}_k = \frac{\hat{p}^2}{2m}$ is the kinetic energy. By Jensen's inequality,  $\langle\hat{E}_k^{-2}\rangle \geq \langle \hat{E}_k\rangle^{-2}$. Hence, Eq. (\ref{ineqq}) implies that
\begin{eqnarray}
\langle \hat{E}_k \rangle \Delta t > \frac{1}{4} \label{ineqq2}
\end{eqnarray}
Eq. (\ref{ineqq2}) for Kijowski's POVM was derived in Ref. \cite{AnSav17}, in reference to an earlier inequality---without the factor $\frac{1}{4}$---of Ref. \cite{BSPME}. We conjecture that the  greatest lower  bound $C$ to $\langle \hat{E}_k \rangle \Delta t$ is greater than $\frac{1}{4}$--- our best guess is that $C = \frac{1}{2}$---but we have not found a proof.

Inequalities analogous to   (\ref{ineqq2}), but with a different physical interpretation, have appeared before.
 Peierls and Landau argued that  particles can be localized only with uncertainty $\Delta x \gtrsim \langle E\rangle^{-1}$ \cite{LP31}.  Margolus and Levitin proved that $\langle \hat{H} \rangle \tau > \frac{\pi}{2}$,  where $\tau$ is the time it takes a quantum  system to evolve
between two orthogonal states \cite{MaLe98}.

 The second term in the r.h.s. of Eq. (\ref{ineqq}) is greater than $\frac{m^4}{4} \langle \epsilon_p^{-6}\rangle$. Hence, Eq. (\ref{ineqq}) implies the weaker inequality
 \begin{eqnarray}
  (\Delta t)^2 \geq \frac{1}{4 (\Delta H)^2} +\frac{m^4}{4} \langle\hat{H}^{-6}\rangle \label{ineqq3}
 \end{eqnarray}
 that is sharper in the ultra-relativistic limit. Using  Jensen's inequality on (\ref{ineqq3}), we find
\begin{eqnarray}
\langle \hat{H} \rangle^3 \Delta t > \frac{m^2}{2}.  \label{ineqq3b}
\end{eqnarray}

We estimate the relative size of the two terms in the r.h.s. of Eq. (\ref{ineqq}), by evaluating them for a two-parameter family of probability densities for the  dimensionless kinetic energy $\xi = (\epsilon_p/m - 1)$,
\begin{eqnarray}
P(\xi) = \sqrt{\frac{\xi_0^3}{2 \pi \xi^3}} \exp \left[ - \frac{\xi_0(\xi -\xi_0)^2}{2 \sigma_{\xi}^2 \xi}\right]. \label{invgaus}
\end{eqnarray}
Eq. (\ref{invgaus}) defines an inverse Gaussian distribution \cite{invgaus} for $\xi \geq 0$, with mean $\xi_0$ and variance $\sigma_{\xi}^2$. It has finite moments $\langle \xi^{n}\rangle$, for all integers $n$ (including negative ones).

In the non-relativistic limit, Eq. (\ref{ineqq}) gives
\begin{eqnarray}
(\Delta t)^2 \geq \frac{1}{4m^2 \xi_0^2} \left[ b + \frac{1}{4}\left(1 +\frac{3}{b}+ \frac{3}{b^2}\right)\right], \label{uncnonr}
\end{eqnarray}
where $b = (\xi_0/\sigma_{\xi})^2$. The term $(\Delta H)^{-2}$ is proportional to $b$, and thus dominates for almost monochromatic states, where $b \rightarrow \infty$. The other term dominates for states with a large momentum spread, $b \rightarrow 0$. Minimizing the r.h.s. of Eq. (\ref{uncnonr}) with respect to $b$ gives a variational estimate to the greatest lower bound of $\langle \hat{E}_k\rangle \Delta t$. We obtain $\langle \hat{E}_k\rangle \Delta t \gtrsim 0.8$.

For the ultra-relativistic limit, where $\hat{H} \simeq \hat{E}_k$, we use (\ref{ineqq3}) to obtain
\begin{eqnarray}
(\Delta t)^2 \geq \frac{1}{4m^2 \xi_0^2} \left[ b + \frac{1}{\xi_0^4}[1+ 21 q(b^{-1})]\right] \label{unceroo}
\end{eqnarray}
 where $q(x) = x +10x^2+60x^3 +225 x^4 +495x^5+495x^6$. For $\xi_0 >> 1$, the dominant contribution to $q(x)$ is the $x^6$ term. Then, minimization with respect to $b$ gives
 \begin{eqnarray}
 \langle H \rangle^{9/7}  \Delta t \gtrsim 1.7 m^{2/7}.
 \end{eqnarray}

\section{Conclusions}
In this paper, we described the localization of relativistic particles in terms of time-of-arrival observables. The latter are defined in terms of the interaction between  the associated quantum field
and the measuring apparatus.  The apparatus is described by QFT at the microscopic level; however, its records are macroscopic and they are expressed in terms of classical spacetime coordinates $X$.  We derived a probability density $P(X)$ for particle detection that is a genuine density with respect to $X$ and applies to any QFT.

Then, we specialized to the simplest case of a neutral scalar field, hence, spinless particles. We obtained a family of time-of-arrival POVMs for the relativistic particle that differ on  the way that the apparatus effects localization. We found that there is a unique POVM that leads to maximal localization. This POVM  is related to the Newton-Wigner position operator, and it reduces to Kijowski's POVM in the non-relativistic limit. Finally, we  derived an uncertainty relation for the time-of-arrival that provides a lower bound to the spacetime localization of relativistic particles.

 The methods developed in this paper can be straightforwardly applied to particles of other spin, including photons, gravitons and neutrinos. We expect that their localization properties are more elaborate,  because  spin strongly affects  the localization operator. Our results  can also be adapted for studying localization in interacting theories (e.g., QCD), as the only input to the analysis is a QFT two-point function.

The time-of-arrival probabilities derived here are expressed in terms of causal QFT correlation functions. For this reason, we believe that their description of particle localization is consistent with relativistic causality. To prove this, we must study  signal propagation in set-ups that involve multiple measurements, for example, by generalizing the analysis of Ref. \cite{AnSav17} to relativistic systems.


\section*{Acknowledgements}
C. A acknowledges support by Grant No. E611 from the Research Committee of the University of Patras via the ”K. Karatheodoris” program.


\begin{appendix}

\section{The QTP probability formula}

In this section, we present a derivation of  the probability formula (\ref{ampl6}). We work in the set-up described in Sec. 2.1, and employ the same notation.

First, we
construct
 the probability amplitude $| \psi; \lambda, [t_1, t_2] \rangle$ that, given an initial ($t=0$) state $|\psi_0\rangle \in {\cal H}_-$, a transition occurs during the time interval $[t_1, t_2]$ and a value $\lambda$ for the pointer variable is obtained
 for
 some observable.

For a vanishingly small time interval, $t_1 = t$ and $t_2 = t + \delta t$, we  keep only leading-order terms with respect to $\delta t$. At times prior to $t$, the state
lies in ${\cal H}_-$. This is taken into account by evolving the initial state $|\psi_0 \rangle$ with the restricted propagator  $\hat{S}_t =  \lim_{N
\rightarrow \infty} (\hat{Q}e^{-i\hat{H} t/N} \hat{Q})^N$
 in ${\cal H}_-$. Since the transition occurs at some instant within the time interval $[t, t+\delta t]$, there is no constraint in the propagation from $t$ to $t + \delta t$; it  is implemented by  the unrestricted evolution operator   $e^{-i \hat{H} \delta t} \simeq 1 - i \delta t \hat{H}$. At time $t + \delta t$, the event corresponding to $\hat{\Pi}(\lambda)$ is recorded, so the
amplitude is transformed by the action of $\sqrt{\hat{\Pi}({\lambda})}$. For times greater than $t + \delta t$, there is no constraint, so the amplitude
evolves as $e^{-i \hat{H} (T-t)}$  until some final moment $T$.

At the limit of small $\delta t$, the successive operations above yield
\begin{eqnarray}
|\psi_0; \lambda, [t, t+ \delta t] \rangle =  - i \, \delta t \, \,e^{-i\hat{H}(T - t)} \sqrt{\hat{\Pi}}(\lambda) \hat{H} \hat{S}_t |\psi_0
\rangle. \label{amp1}
\end{eqnarray}

 The amplitude $|\psi_0; \lambda, [t, t + \delta t] \rangle$ is proportional to $\delta t$. Therefore, it defines  a {\em density} with respect to time, $|\psi_0;  \lambda, t \rangle := \lim_{\delta t \rightarrow 0}
\frac{1}{\delta t} | \psi_0; \lambda, [t, t + \delta t] \rangle$.
 Eq. (\ref{amp1}) implies that

\begin{eqnarray}
 |\psi_0;  \lambda, t \rangle = - i
e^{- i \hat{H} T} \hat{C}(\lambda, t) |\psi_0 \rangle, \label{amp2}
\end{eqnarray}
where   the {\em class operator} $\hat{C}(\lambda, t)$ is
 \begin{eqnarray}
  \hat{C}(\lambda, t) := e^{i \hat{H}t} \sqrt{\hat{\Pi}}(\lambda)
\hat{H} \hat{S}_t. \label{class}
\end{eqnarray}

 Since the amplitude $|\psi_0;  \lambda, t \rangle $ is a density
with respect to   $t$,  integration over $t$ is well defined.   Thus, the total amplitude that the transition occurred at {\em some moment} within a time interval $[t_1, t_2]$ is

\begin{eqnarray}
| \psi; \lambda, [t_1, t_2] \rangle = - i e^{- i \hat{H}T} \int_{t_1}^{t_2} d t \hat{C}(\lambda, t) |\psi_0 \rangle. \label{ampl5}
\end{eqnarray}

Here, we   consider  Hamiltonians of the form   $\hat{H} = \hat{H_0} + \hat{H_I}$, where
$[\hat{H}_0, \hat{P}] = 0$, and $H_I$ a perturbing interaction. To leading order in the perturbation,
$ \hat{C}(\lambda, t) = e^{i \hat{H}_0t} \sqrt{\hat{\Pi}}(\lambda) \hat{H}_I e^{-i \hat{H}_0t}$.

By Born's rule, the squared modulus of the amplitude  Eq. (\ref{amp2}) should define
the probability $P (\lambda, [t_1, t_2])\/$ that at some time in the interval $[t_1, t_2]$ a detection with outcome $\lambda$ occurred,
\begin{eqnarray}
 P(\lambda, [t_1, t_2])
:= \langle \psi; \lambda, [t_1, t_2] | \psi; \lambda, [t_1, t_2] \rangle =   \int_{t_1}^{t_2} \,  dt \, \int_{t_1}^{t_2} dt' Tr\left(\hat{C}(\lambda, t)
\hat{\rho_0}\hat{C}^{\dagger}(\lambda, t')\right)  , \label{prob1}
\end{eqnarray}
where $\hat{\rho}_0 = |\psi_0\rangle \langle \psi_0|$.

However, the quantities $P(\lambda, [t_1, t_2])$ do not define a probability measure with respect to time $t$, because they do not satisfy the Kolmogorov additivity axiom of probabilities. To see this, consider the
probability corresponding to an
 interval $[t_1, t_3] = [t_1, t_2] \cup [t_2,
t_3]$,
 \begin{eqnarray}
 P(\lambda, [t_1, t_3]) = P(\lambda, [t_1, t_2]) + P(\lambda, [t_2, t_3]) + 2 Re \left[ \int_{t_1}^{t_2} \,  dt \, \int_{t_2}^{t_3} dt' Tr\left(\hat{C}(\lambda, t)
\hat{\rho_0}\hat{C}^{\dagger}(\lambda, t')\right)\right]. \label{add}
\end{eqnarray}

The Kolmogorov additivity condition $P(\lambda, [t_1, t_3]) = P(\lambda, [t_1, t_2]) + P(\lambda, [t_2, t_3])$ fails, unless

\begin{eqnarray}
 2 Re \left[ \int_{t_1}^{t_2} \,  dt \, \int_{t_2}^{t_3} dt' Tr\left(\hat{C}(\lambda, t) \hat{\rho_0}\hat{C}^{\dagger}(\lambda, t')\right)\right] = 0 \label{decond}
 \end{eqnarray}
 In the consistent/decoherent histories framework, Eq. (\ref{decond}) is referred to as the
 {\em consistency condition} \cite{Omn1, Omn2, Gri}. It is the minimal condition necessary for defining a consistent probability measure for histories.

   Eq. (\ref{decond}) does not hold for generic choices of $t_1, t_2$ and $t_3$. However, we expect that it holds given a
   sufficient degree of coarse-graining. That is, we assume that there exists a   time-scale $\sigma_T$, such that the non-additive terms in Eq. (\ref{add}) are strongly
   suppressed if   $ |t_2 - t_1| >  \sigma_T$ and $|t_3 - t_2| >  \sigma_T$. This is a natural assumption  for a system that involves a macroscopic component such as a measuring apparatus  \cite{GeHa, hartlelo}. Then, Eq. (\ref{prob1}) defines a probability measure when restricted to intervals of size  larger than $\sigma_T$.

   The probabilities with respect to  $\lambda$ are consistently defined, if
   \begin{eqnarray}
   \langle \psi; \lambda, [t_1, t_2] | \psi; \lambda', [t_1, t_2] \rangle \simeq  \delta_{\lambda \lambda'} \langle \psi; \lambda, [t_1, t_2] | \psi; \lambda, [t_1, t_2] \rangle, \label{decla}
   \end{eqnarray}
for $|t_2 - t_1| > \sigma_T$, and for $\lambda$ a discrete variable. This is a constraint on the positive operators $\hat{\Pi}(\lambda)$ that can represent a record $\lambda$.    Equivalently, Eq. (\ref{decla}) can be viewed as a {\em definition} of what it means for a class of positive operators $\hat{\Pi}(\lambda)$ to represent a macroscopic record of observation. For continuous variables $\lambda$, the analogous condition is Eq. (\ref{deccond}).

We   define the time-of-transition probabilities by smearing the amplitudes
 Eq. (\ref{amp2}) at the  time-scale  $\sigma_T$ rather than using  sharp time-intervals, as in Eq. (\ref{prob1}). Then, the time-of-transition  probabilities are expressed in terms of
 densities
of a continuous time variable. To this end, we introduce a family of sampling functions $f(s)$,  localized around $s = 0$ with width $\sigma_T$, and normalized so that $\int ds f(s) = 1$.

We  define the smeared amplitude $|\psi_0; \lambda, t\rangle_{\sigma_T}$ that is localized
 around the time $t$, as
\begin{eqnarray}
|\psi_0; \lambda, t\rangle_{\sigma_T} := \int ds \sqrt{f(t-s)} |\psi_0; \lambda, s \rangle =  \int ds \sqrt{f(t-s)} \hat{C}(\lambda, s) |\psi_0 \rangle, \label{smearing}
\end{eqnarray}
 Then, the modulus-
squared amplitudes  $P(\lambda, t) = {}_{\sigma_T}\langle \psi_0; \lambda, t|\psi_0; \lambda, t\rangle_{\sigma_T} $
give  Eq. (\ref{ampl6}).

\end{appendix}

 \end{document}